\documentclass[sn-mathphys-num]{sn-jnl}

\usepackage{graphicx}%
\usepackage{multirow}%
\usepackage{amsmath,amssymb,amsfonts}%
\usepackage{amsthm}%
\usepackage{mathrsfs}%
\usepackage[title]{appendix}%
\usepackage{xcolor}%
\usepackage{textcomp}%
\usepackage{manyfoot}%
\usepackage{booktabs}%
\usepackage{algorithm}%
\usepackage{algorithmicx}%
\usepackage{algpseudocode}%
\usepackage{listings}%
\usepackage[numbers]{natbib}

\theoremstyle{thmstyleone}%
\theoremstyle{thmstyletwo}%

\theoremstyle{thmstylethree}%

\begin{document}

\title[Identification of $>$40 gravitationally magnified stars in a galaxy at redshift of 0.725]{Identification of $>$40 gravitationally magnified stars in a galaxy at redshift of 0.725}

\author*[1]{\fnm{Yoshinobu} \sur{Fudamoto}}\email{yoshinobu.fudamoto@gmail.com}
\author[2,3]{\fnm{Fengwu} \sur{Sun}}
\author[4]{\fnm{Jose M.} \sur{Diego}}
\author[5]{\fnm{Liang} \sur{Dai}}
\author[1,6]{\fnm{Masamune} \sur{Oguri}}
\author[7]{\fnm{Adi} \sur{Zitrin}}
\author[8,9]{\fnm{Erik} \sur{Zackrisson}}
\author[10,11,12,13]{\fnm{Mathilde} \sur{Jauzac}}
\author[10,11]{\fnm{David J.} \sur{Lagattuta}}
\author[2]{\fnm{Eiichi} \sur{Egami}}
\author[14]{\fnm{Edoardo} \sur{Iani}}
\author[15]{\fnm{Rogier A.} \sur{Windhorst}}
\author[1]{\fnm{Katsuya T.} \sur{Abe}}
\author[16,17,18]{\fnm{Franz Erik} \sur{Bauer}}
\author[19]{\fnm{Fuyan} \sur{Bian}}
\author[20]{\fnm{Rachana} \sur{Bhatawdekar}}
\author[21,22,23]{\fnm{Thomas J.} \sur{Broadhurst}}
\author[24]{\fnm{Zheng} \sur{Cai}}
\author[25]{\fnm{Chian-Chou} \sur{Chen}}
\author[26]{\fnm{Wenlei} \sur{Chen}}
\author[27]{\fnm{Seth H.} \sur{Cohen}}
\author[28]{\fnm{Christopher J.} \sur{Conselice}}
\author[29,30]{\fnm{Daniel} \sur{Espada}}
\author[31]{\fnm{Nicholas} \sur{Foo}}
\author[32]{\fnm{Brenda L.} \sur{Frye}}
\author[33]{\fnm{Seiji} \sur{Fujimoto}}
\author[34]{\fnm{Lukas J.} \sur{Furtak}}
\author[34]{\fnm{Miriam} \sur{Golubchik}}
\author[35]{\fnm{Tiger Yu-Yang} \sur{Hsiao}}
\author[36]{\fnm{Jean-Baptiste} \sur{Jolly}}
\author[1,37]{\fnm{Hiroki} \sur{Kawai}}
\author[38]{\fnm{Patrick L.} \sur{Kelly}}
\author[39]{\fnm{Anton M.} \sur{Koekemoer}}
\author[40,41]{\fnm{Kotaro} \sur{Kohno}}
\author[14]{\fnm{Vasily} \sur{Kokorev}}
\author[24]{\fnm{Mingyu} \sur{Li}}
\author[24]{\fnm{Zihao} \sur{Li}}
\author[24,2]{\fnm{Xiaojing} \sur{Lin}}
\author[42,43,44]{\fnm{Georgios E.} \sur{Magdis}}
\author[7]{\fnm{Ashish K.} \sur{Meena}}
\author[45,10,11]{\fnm{Anna} \sur{Niemiec}}
\author[46]{\fnm{Armin} \sur{Nabizadeh}}
\author[47]{\fnm{Johan} \sur{Richard}}
\author[48,44]{\fnm{Charles L.} \sur{Steinhardt}}
\author[24]{\fnm{Yunjing} \sur{Wu}}
\author[2]{\fnm{Yongda} \sur{Zhu}}
\author[24,49]{\fnm{Siwei} \sur{Zou}}

\affil*[1]{\orgdiv{Center for Frontier Science}, \orgname{Chiba University}, \orgaddress{1-33 Yayoi-cho, Inage-ku, Chiba 263-8522, Japan}}
\affil[2]{\orgdiv{Steward Observatory}, \orgname{University of Arizona}, \orgaddress{933 N. Cherry Ave., Tucson, AZ 85721, USA}}
\affil[3]{\orgdiv{Center for Astrophysics}, \orgname{ Harvard \& Smithsonian}, \orgaddress{60 Garden St., Cambridge, MA 02138, USA}}
\affil[4]{\orgdiv{Instituto de F{\'i}sica de Cantabria (CSIC-UC).}, \orgaddress{Avda. Los Castros 39005 Santander, Spain}}
\affil[5]{Department of Physics, University of California, 366 Physics North MC 7300, Berkeley, CA. 94720, USA}
\affil[6]{Department of Physics, Graduate School of Science, Chiba University, Chiba 263-8522, Japan}
\affil[7]{Department of Physics, Ben-Gurion University of the Negev, P.O. Box 653, Be'er-Sheva 84105, Israel}
\affil[8]{Observational Astrophysics, Department of Physics and Astronomy, Uppsala University, Box 516, SE-751 20 Uppsala, Sweden}
\affil[9]{Swedish Collegium for Advanced Study, Linneanum, Thunbergsv\"a{}gen 2, SE-752 38 Uppsala, Sweden}
\affil[10]{Centre for Extragalactic Astronomy, Durham University, South Road, Durham DH1 3LE, UK}
\affil[11]{Institute for Computational Cosmology, Durham University, South Road, Durham DH1 3LE, UK}
\affil[12]{Astrophysics Research Centre, University of KwaZulu-Natal, Westville Campus, Durban 4041, South Africa}
\affil[13]{School of Mathematics, Statistics \& Computer Science, University of KwaZulu-Natal, Westville Campus, Durban 4041, South Africa}
\affil[14]{Kapteyn Astronomical Institute, University of Groningen, P.O. Box 800, 9700 AV Groningen, The Netherlands}
\affil[15]{School of Earth and Space Exploration, Arizona State University, Tempe, AZ 85287-6004, USA}
\affil[16]{Instituto de Astrof{\'{\i}}sica and Centro de Astroingenier{\'{\i}}a, Facultad de F{\'{i}}sica, Pontificia Universidad Cat{\'{o}}lica de Chile, Campus San Joaquín, Av. Vicuña Mackenna 4860, Macul Santiago, Chile, 7820436}
\affil[17]{Millennium Institute of Astrophysics, Nuncio Monse{\~{n}}or S{\'{o}}tero Sanz 100, Of 104, Providencia, Santiago, Chile}
\affil[18]{Space Science Institute, 4750 Walnut Street, Suite 205, Boulder, Colorado 80301}
\affil[19]{European Southern Observatory, Alonso de Córdova 3107, Casilla 19001, Vitacura, Santiago 19, Chile}
\affil[20]{European Space Agency (ESA), European Space Astronomy Centre (ESAC), Camino Bajo del Castillo s/n, 28692 Villanueva de la Cañada, Madrid, Spain}
\affil[21]{Dept. of Physics, University of the Basque Country, Bilbao, Spain}
\affil[22]{DIPC, Donostia International Physics Centre, San Sebastian, Spain}
\affil[23]{Ikerbasque, Basque Foundation for Science, Bilbao, E-48011, Spain}
\affil[24]{Department of Astronomy, Tsinghua University, Beijing 100084, China}
\affil[25]{Academia Sinica Institute of Astronomy and Astrophysics (ASIAA), No. 1, Sec. 4, Roosevelt Road, Taipei 10617, Taiwan}
\affil[26]{Department of Physics, Oklahoma State University, 145 Physical Sciences Bldg, Stillwater, OK 74078, USA}
\affil[27]{School of Earth and Space Exploration, Arizona State University, Tempe, AZ 85287-1404, USA}
\affil[28]{Jodrell Bank Centre for Astrophysics, University of Manchester, Oxford Road, Manchester M13 9PL, UK}
\affil[29]{Departamento de F\'{i}sica Te\'{o}rica y del Cosmos, Campus de Fuentenueva, Edificio Mecenas, Universidad de Granada, E-18071, Granada, Spain}
\affil[30]{Instituto Carlos I de F\'{i}sica Te\'{o}rica y Computacional, Facultad de Ciencias, E-18071, Granada, Spain}
\affil[31]{School of Earth and Space Exploration, Arizona State University, Tempe, AZ 85287-1404, USA}
\affil[32]{Department of Astronomy/Steward Observatory, University of Arizona, 933 N. Cherry Avenue, Tucson, AZ 85721, USA}
\affil[33]{Department of Astronomy, The University of Texas at Austin, Austin, TX 78712, USA}
\affil[34]{Department of Physics, Ben-Gurion University of the Negev, P.O. Box 653, Be'er-Sheva 84105, Israel}
\affil[35]{Center for Astrophysical Sciences, Department of Physics and Astronomy, The Johns Hopkins University, 3400 N Charles St. Baltimore, MD 21218, USA}
\affil[36]{Max-Planck-Institut für Extraterrestrische Physik (MPE), Giessenbachstraße 1, D-85748 Garching, Germany}
\affil[37]{Department of Physics, The University of Tokyo, Bunkyo, Tokyo 113-0033, Japan}
\affil[38]{Minnesota Institute for Astrophysics, University of Minnesota, 116 Church Street SE, Minneapolis, MN 55455, USA}
\affil[39]{Space Telescope Science Institute, 3700 San Martin Drive, Baltimore, MD 21218, USA}
\affil[40]{Institute of Astronomy, Graduate School of Science, The University of Tokyo, 2-21-1 Osawa, Mitaka, Tokyo 181-0015, Japan}
\affil[41]{Research Center for the Early Universe, Graduate School of Science, The University of Tokyo, 7-3-1 Hongo, Bunkyo-ku, Tokyo 113-0033, Japan}
\affil[42]{Cosmic Dawn Center (DAWN), }
\affil[43]{DTU-Space, Technical University of Denmark, Elektrovej 327, 2800, Kgs. Lyngby, Denmark}
\affil[44]{Niels Bohr Institute, University of Copenhagen, Jagtvej 128, 2200, Copenhagen N, Denmark}
\affil[45]{LPNHE, CNRS/IN2P3, Sorbonne Universit{\'e}, Universit{\'e} Paris-Cit{\'e}, Laboratoire de Physique Nucl{\'e}aire et de Hautes {\'E}nergies, F-75005 Paris, France}
\affil[46]{Observational Astrophysics, Department of Physics and Astronomy, Uppsala University, Box 516, SE-751 20 Uppsala, Sweden}
\affil[47]{Univ Lyon, Univ Lyon1, Ens de Lyon, CNRS, Centre de Recherche Astrophysique de Lyon UMR5574, F-69230, Saint-Genis-Laval, France}
\affil[48]{Dark Cosmology Centre, Niels Bohr Institute, University of Copenhagen, Jagtvej 155, København N, DK-2200, Denmark}
\affil[49]{Chinese Academy of Sciences South America Center for Astronomy, National Astronomical Observatories, CAS, Beijing 100101, China}

\abstract{
    Strong gravitational magnifications enable to detect faint background sources, resolve their internal structures, and even identify individual stars in distant galaxies.
    Highly magnified individual stars allow various applications, including studies of stellar populations in distant galaxies and constraining dark matter structures in the lensing plane.
    However, these applications have been hampered by the small number of individual stars observed, as typically one or a few stars are identified from each distant galaxy.
    Here, we report the discovery of more than 40 microlensed stars in a single galaxy behind Abell 370 at redshift of 0.725 when the Universe was half of its current age (dubbed ``the Dragon arc''), using  James Webb Space Telescope ({\it JWST}) observations with the time-domain technique.
    These events are found near the expected lensing critical curves, suggesting that these are magnified stars that appear as transients from intracluster stellar microlenses. Through multi-wavelength photometry, we constrain stellar types and find that many of them are consistent with red giants/supergiants magnified by factors of hundreds.
    This finding reveals an unprecedented high occurrence of microlensing events in the Dragon arc, and proves that {\it JWST}'s time-domain observations open up the possibility of conducting statistical studies of high-redshift stars.
    }

\maketitle

The high magnification afforded by massive galaxy clusters accompanied with microlensing enable us to identify individual stars in distant galaxies\cite{1991ApJ...379...94M,Kelly2018}.
Highly magnified individual stars in distant galaxies allow for a wide range of applications to several astronomy fields, including constraints on compact dark matter\cite{2017ApJ...850...49V,2018PhRvD..97b3518O,2018ApJ...857...25D,VallMuler2024} or self-gravitating micro-structures\cite{Dai2020}, the measurement of the abundance of dark matter subhalos\cite{2018ApJ...867...24D, Dai2020s1226,2024PhRvD.109h3517A,2024ApJ...961..200W}, direct constraints on the stellar populations and initial mass function of distant galaxies\cite{Kelly2018,Han2024}, and direct observations of Population III stars \cite{2018ApJS..234...41W,Schauer2022,2024MNRAS.533.2727Z}.
To provide meaningful constraints, these applications require a statistically sufficient large number of distant stars identified from each lensing field. 
However, from previous observations, typically only one or a few individual stars have been identified from each galaxy\cite{Rodney2018,Kelly2018,Kaurov2019,Kelly2023,Diego2023,Welch2022}, and such applications of individual stars have been so far limited.
With its superb light-collecting power and excellent spatial resolution, {\it JWST} has been expected to change the current situation by significantly increasing the detections of individually lensed stars from each galaxy, which finally will open up various applications of individual star detections.

Individual stars require extremely strong gravitational magnifications of factors of hundreds to thousands to become bright enough to be detectable\cite{Kelly2018}. Such strong gravitational magnifications are possible by combinations of strong lenses induced by massive dark matter in galaxy clusters and microlensing events by intervening compact masses such as intracluster stars.
Since microlensing events are usually observed as variable sources, we typically need time-domain observations for extragalactic survey fields to search individual stars.
Here, we report the results of serendipitously obtained {\em JWST}'s time-domain observations of a strongly lensed galaxies.

Our target is a strongly lensed star-forming galaxy at $z=0.725$, behind the galaxy cluster Abell 370, which appears as a giant lensed arc also known as ``the Dragon arc''.
{\em JWST} NIRCam multi-wavelength images of the Dragon arc are obtained over two epochs. One is during {\em JWST} Cycle-1 in December 2022 and another is during Cycle-2 in December 2023. Both observations serendipitously use filters covering similar observing wavelength ranges within which $2\,{\rm \mu m}$ and $4.1\,{\rm \mu m}$ images are commonly observed. The repeat observations allow us to perform a time-domain observation of the Dragon arc.

Using $2\,{\rm \mu m}$ and $4.1\,{\rm \mu m}$ images obtained across the two epochs, we searched transient events that show up only in one of the epochs.
For searching transients, we use the $2\,{\rm \mu m}$ images as the detection images as the $2\,{\rm \mu m}$ images are deeper by $\sim1\,{\rm mag}$ and have higher spatial resolution than the $4.1\,{\rm \mu m}$ images. 
The two epoch observations of the $2\,{\rm \mu m}$ wavelength images apparently show a large number of compact transients across the Dragon arc (bottom panels of Figure 1).
To more precisely identify fainter transients, we created a differential image by subtracting each epoch's $2\,{\rm \mu m}$ observation (see Methods for more details).
Thanks to the high spatial resolution and high sensitivity of each $2\,{\rm \mu m}$ image with the identical position angle of the telescope, we obtained a clean differential $2\,{\rm \mu m}$ image which shows a large number of transient events appearing as bright positive and negative detections in the differential image (Figure 2).
We used two complementary source-finding algorithms (\texttt{DAOFIND}\cite{Stetson1987} and \texttt{Sextractor}\cite{Bertin1996}) to efficiently detect compact and crowded transients from the $2\,{\rm \mu m}$ differential image (see Methods for more details).
In total, we identified 45 bright and securely detected transients with the signal to noise ratio (SNR)  of greater than $5$ across  the Dragon arc.

After careful investigations, we find one transient $d_{\perp}\sim1^{\prime\prime}.8$ away from the Dragon arc ({\em event $\alpha$} in Figure 2) that could likely be a non-lensed type of transients (e.g., the late stage of a supernova), where $d_{\perp}$ represents the distance from the closest position to the Dragon arc (see Methods).
However, finding more than $40$ supernovae simultaneously from a single galaxy is improbable even in this faint limit\cite{DeCoursey2024}.
Also, transients seen in the $2\,{\rm \mu m}$ images are located around the expected positions of critical curve within the Dragon arc, while each lens model has slightly different predictions for the positions of critical curves due to uncertainties of the models. Thus these transients should be strongly magnified with factors of $\sim 10$ to $\sim6000$ that are estimated from various lensing models.
As they appear as transients typical for microlensing events, from their expected high lensing magnifications, and their point source morphology in the image plane, we conclude that these transients represent microlensed stars in the Dragon arc.

To study basic properties of the lensed individual stars, we create a color magnitude diagram using photometry from differential images of $2\,{\rm \mu m}$ and $4.1\,{\rm \mu m}$.
To study the color magnitude diagram, we limit our sample only to F410M bright sources, namely F410M detection of $>5\,\sigma$ as F410M images are shallower than F200W image and individual detection of the lensed stars are limited (see Methods).
With these criteria, we analyze the color magnitude diagram for 8 micro-lensed stars. 
We find that these stars have rest-frame F200W - F410M colors consistent with stars having surface temperatures of $\sim3000{\rm K}$ -- $4000\,{\rm K}$\cite{Lejeune1997} (Figure 3). 
Such low surface temperature indicate that these stars are either low mass main-sequence stars or red giants/supergiants with apparent luminosity of $\mu\,L\sim10^{7-8}\,{\rm L_{\odot}}$ where $\mu$ represents the lensing  magnification factor.
As these microlensing events typically have magnification factors of a few thousand, these results indicate that the F200W and F410M detected stars are red supergiants with intrinsic liminosities of $L_{\ast}\sim10^{5}\,{\rm L_{\odot}}$ receiving gravitational magnifications of factors $\mu= 100{\rm s}$ -- $1000$.

Excluding one likely non-lensing transients, the finding of 44 microlensed stars in a single high-redshift galaxy far surpasses previous records\cite{Kelly2018,Rodney2018}.
The findings of the large number of highly magnified individual stars demonstrate the unique power of sensitive time-domain observations in near-IR wavelengths using {\it JWST} as well as the high occurrence of microlensing events in the Dragon arc.
Future {\it JWST} time-domain observations of the Dragon arc as well as other similar strongly lensed galaxies at high redshift will allow us to obtain large number of individual stars detected from galaxies in cosmological distances, which will provide us with essential information about stellar populations in high-redshift galaxies.
A full SED analysis of magnified stars will be enabled by another imaging visit using the same filters in future observations.
The discovery of the large number of highly magnified individual stars show that {\it JWST} has open up the new possibility of conducting statistical studies of high-redshift stars and subgalactic scale perturbations in the lensing field.
\break
\begin{figure}
    \centering
    \includegraphics[width=1.0\textwidth]{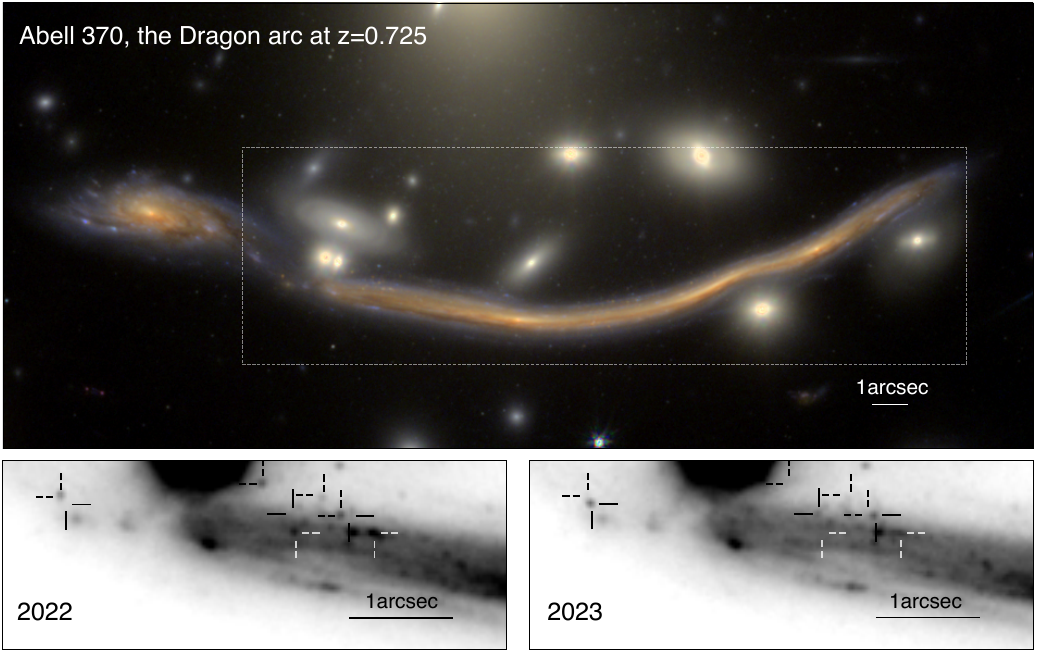}
    \caption{
    \textbf{Figure 1 | JWST observations of distant stars seen as bright transients in the Dragon arc.}
    \textit{ \textbf{Upper panel}}: A false-color image of the entire ``the Dragon arc'' behind Abell 370 cluster \cite{Soucial1987b,Lynds1989}, using {\it JWST} filters F090W, F150W, and F200W. North is up, East to the left, and a reference angular scale of $1^{\prime\prime}$ is shown by the solid horizontal bar at the bottom right corner.
    The dashed white rectangle shows the region of interest further analyzed in Figure 2.
    \textit{\textbf{Lower panels}}: F200W zoom in on a part of the Dragon arc in the 2022 image (left panel) and in the effective F200W in 2023 (right panel). The effective F200W image in 2023 was made using F182M and F210M images (see Methods). Examples of the apparently bright microlensing events are indicated, where dashed half-crosses show bright sources seen only in 2022 data and solid half-crosses show sources only in 2023. Many additional microlensing events exist, but are only visible in the differential image in Figure 2. Horizontal bars in the lower right corners show $1^{\prime\prime}$ scales.}
    \label{fig:dragon}
    \vspace{5cm}
\end{figure}

\begin{figure}[!t]
    \centering
    \includegraphics[width=1.0\textwidth]{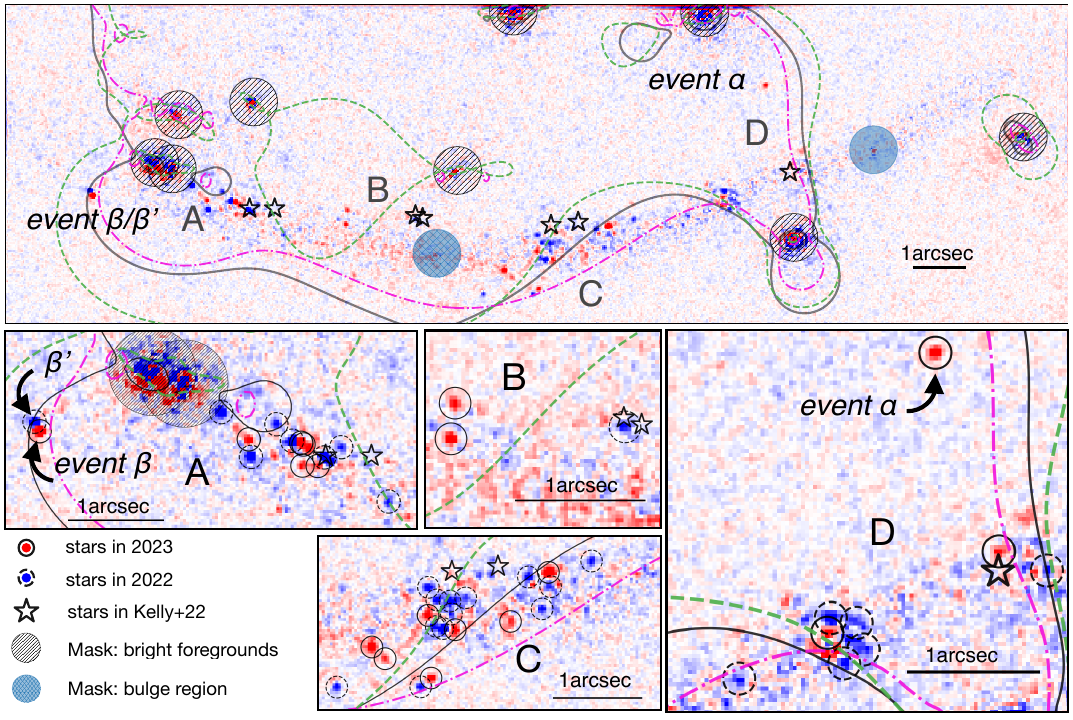}
    \caption{
    \textbf{Figure 2 $|$ $2\,{\rm \mu m}$ differential images of the Dragon arc between the 2022 and 2023 epochs.}
    \textit{\textbf{Top Panel}}: The entire $2\,{\rm \mu m}$ differential images. Positive signals (red) show objects that appear only in 2023, while negative signals (blue) show sources only seen in 2022. 
    Contours show critical curves for $z=0.725$ estimated from several programs: 
    the solid black critical curve is derived with the \texttt{WSLAP+} code \cite{Diego2007,2018MNRAS.473.4279D}, 
    the dashed green lines are from the \texttt{lenstool} software \cite{2009MNRAS.395.1319J,2023MNRAS.524.2883N}, 
    and the dash-dotted purple lines are from the Light-Traces-Mass (LTM) method \cite{Zitrin2009}. 
    Star symbols indicate locations of previous {\it HST} detections of microlensing events \cite{2022arXiv221102670K}.
    \textit{\textbf{Lower Panels}}: Zoom-ins of crowded transient regions indicated with A, B, C, and D letters in the upper panel. Circles ($r=0^{\prime\prime}.12$) show detected microlensed events. Solid circles show lensed stars significantly detected in 2023 epoch, while dashed circles indicate stars detected events in 2022 epoch. Hatched circles show the masked regions that are used to avoid contamination from bright residuals. In total, 44 microlensed stars and 1 likely Supernova are significantly detected.
    }
    \label{fig:diff-image}
    \vspace{5cm}
\end{figure}

\begin{figure}
    \centering
    \includegraphics[width=0.75\columnwidth]{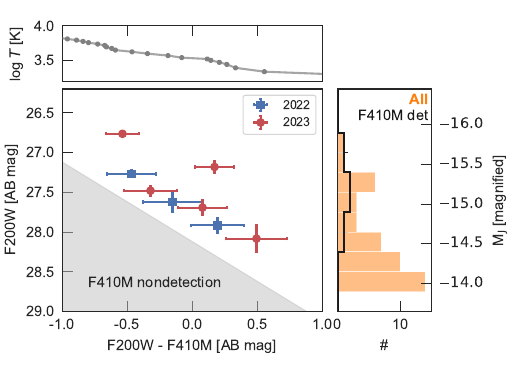}
    \caption{
    \textbf{Figure 3 $|$ $2\,{\rm \mu m}$ Color-magnitude diagram of F410M-detected transients.}
    Square symbols show events in 2022 and circular symbols show events in 2023.
    Filled space shows the F410M detection limit. Except for {\it event} $\alpha$, $\beta$, $0.4\,{\rm mag}$ of dust attenuation corrections are applied for all F200W - F410M colors estimated from measurements of \cite{Patricio2018}.
    {Error bars represent the estimated standard deviations of measured colors. The gray area shows the color space inaccessible among our data due to the sensitivity of the F410M difference image.}
    Solid line in the upper panel shows typical F200W - F410M color of different temperature giant stars redshifted at $z=0.725$ \cite{Lejeune1997}.
    Right panel shows a histogram of detected F200W magnitude and absolute magnitude at the rest-frame J-band.
    The observed J-band absolute magnitude of these sources show that these stars are strongly magnified by $\mu\gtrsim100{\rm s} - 1000$ assuming they are red giants or asymptotic giant stars \cite{Madore2022ApJ}.
    }
    \label{fig:color}
    \vspace{15cm}
\end{figure}

\break
\noindent\textbf{\large Methods}
\section*{Cosmology}
Throughout this paper we assume a concordance cosmology with $\Omega_{m}=0.3$, $\Omega_{\Lambda}=0.7$, $\mathrm{h}=0.7$.

\section*{Observations and Reduction of {\it JWST} Data} \label{sec:obs}
We used {\it JWST} NIRCam imaging \cite{Rieke23} observations from two programs: Cycle-1 GTO-1208 (the CAnadian NIRISS Unbiased Cluster Survey, CANUCS, PI: C. Willot \cite{Willot22}) and Cycle-2 GO-3538 (PI: E. Iani) which are publicly available.
These two programs targeted Abell 370 cluster at $z=0.375$ using multiple NIRCam filters.
Observations were executed in December 2022 and December 2023, respectively, i.e., separated in a time period of $\sim1$ year. Abell 370 field includes a giant arc of a star-forming galaxy at $z=0.725$ also known as the ``Dragon''\cite{Soucial1987b,Soucail1987,Lynds1989}, which is the target of this study.

GTO-1208 CANUCS obtained images in filters of F090W, F115W, F150W, F200W, F277W, F356W,  F410M, F444W, each with integration times of $\sim6400\,{\rm s}$. GO-3538 performed NIRCam wide-field slitless spectroscopy and direct imaging in filters of F300M, F335M, F410M, and F460M, accompanied with short-wavelength imaging using F182M ($\sim19400\,{\rm s}$) and F210M ($\sim19090\,{\rm s}$) filters. Direct imaging for long wavelength filters were also taken with integration time of 2770--3092\,${\rm s}$.

All observations are calibrated using the standard {\it JWST} pipeline \cite{Bushouse2023} version v1.11.2, with reference file \texttt{jwst\_1188.pmap}. We also included customized routines to remove well-known artifacts, including the subtraction of 1/f noise stripes in both row and column directions, template-based wisp subtraction in NIRCam short-wavelength detectors, hot-pixel masking in long-wavelength detectors, and manual masking of the persistence from bright objects \cite{Rigby22,Rieke23,Rieke23b}.
The astrometry of each image was carefully corrected using a combined catalog of HST sources detected with HFF/BUFFALO images \cite{Steinhardt20} and DESI Legacy Imaging Survey catalog\cite{Dey19}, both registered to \textit{Gaia}\cite{Gaia2018}. As a result, each image is well aligned, with an absolute astrometric error of $<0^{\prime\prime}.03$, and an internal RMS astrometric error of $\approx 0^{\prime\prime}.004$ for all $>5\sigma$ sources.
The final mosaicked images were drizzled with \texttt{pix\_frac=1.0} and a pixel size of $0^{\prime\prime}.03$.

\section*{Identifications of microlensing events}
\label{sec:identification}

Our target is a strongly lensed star-forming galaxy, ``the Dragon arc'' at $z=0.725$ behind Abell 370.
This highly lensed galaxy was originally known as ``the Giant Arc''\cite{Soucial1987b,Soucail1987,Lynds1989}. Later, ``Dragon arc'' was labeled during one of image release of the Hubble Space Telescope's Advanced Camera for Surveys (ACS) in 2009.

We first investigated the Dragon arc by directly comparing the F200W (2022) and F182M (2023) images. where we find that several bright sources are only seen in one of the epochs, representing potential transient events across  the Dragon arc. Then, to more precisely identify fainter transient events, we subtracted a combined F182M + F210M image (2023) from the F200W image (2022). 
We use $\lambda_{\rm obs}=2\,{\rm \mu m}$ images to identify transients instead of the repeated F410M filter because the $2\,{\rm \mu m}$ data are deeper by $\sim1$\,mag and higher spatial resolution.
To make the 2023 epoch F182M + F210M image (i.e., ``effective'' F200W image), we first performed point spread function (PSF) homogenization for the F182M and F200W images to match their PSFs to that of F210M.
The PSF for each image was made using \texttt{WebbPSF} \cite{Perrin2014} applying the wavefront measurements for the closest date of each observation.
Convolution kernels were produced using \texttt{pypher} \cite{Boucaud_Convolution_kernels_for_2016} by applying a regularization factor of $0.0001$.
The kernels were convolved using the \texttt{convolve2d} function of the \texttt{scipy.signal} submodule \cite{2020SciPy-NMeth}.
Using the PSF-homogenized F182M and F210M images, we created an effective F200W image by linearly interpolating fluxes at the pivot wavelength of each filter.
We then verified that the effective F200W-2023 image provides consistent fluxes with the F200W-2022 image by performing aperture flux measurements of bright ($m_{\rm F200W}>23.2\,{\rm mag}$) sources in both images. We find the difference of the fluxes are $0.4^{+0.6}_{-1.8}\,\%$.
The differential image was then created by subtracting the effective F200W-2023 image from the PSF-homogenized F200W-2022 image.

Finally, we subtracted the global background of the differential image. This is to remove any large-scale background in the differential image. The systemic large-scale background is made by the differences in background (sky and intracluster light) subtraction during the production of mosaicked images.
As a result, a clean differential image was created with the pixel distribution of $\sim 0.0\pm0.8\,{\rm nJy}$.
The differential image clearly shows significant positive and negative peaks across the Dragon arc, indicating a large number of transient events observed over the two epochs (Figure 2).

We used \texttt{DAOFIND} \cite{Stetson1987} incorporated in the \texttt{photutils.detection} submodule and \texttt{SExtractor} \cite{Bertin1996} to detect microlensing transients from the differential image.
These two source-finding algorithms are complementary in a way that \texttt{DAOFIND} efficiently detects circular Gaussian sources in crowded regions while \texttt{Sextractor} can detect less circular sources that can also be found in the crowded regions in the differential image.

For the \texttt{Sextractor} run, to detect compact sources, we applied a minimum number of connected pixels of four, each of them with flux $>2.5\,\sigma$ above the background standard deviation by setting \texttt{DETECT\_MINAREA $=4$} and \texttt{DETECT\_THRESH} $=2.5$ (i.e., $>5\,\sigma$ detection).
To deblend crowded sources, we applied a contrast parameter of \texttt{DEBLEND\_MINCONT $=0.0005$}.
With these settings, we run \texttt{Sextractor} for positive and negative instances of the differential image, corresponding to transient events in 2023 and 2022, respectively.
For \texttt{DAOFIND} run, we used a $5\sigma$ threshold of $0.038\,{\rm MJy/sr}$ ($0.7$\,nJy\,pix$^{-1}$) and a kernel full width at half maximum of $2.5\,{\rm pixels}$. Similarly, \texttt{DAOFIND} was used for positive and negative instances of the differential image.

For both of the source detections, we masked regions close to bright sources (e.g., bright cluster member galaxies and bulges of the Dragon arc) exist in the original F200W image. 
This is to avoid the contamination from residuals caused by minor PSF differences between 2022 and 2023, which is apparent in the differential image.
All masks have a fixed radius of $r=0^{\prime\prime}.45$ and are centered on the brightest pixel of each galaxy and bulge. To be conservative, the radius is manually determined using the brightest and largest residuals located in the north west and south west part of the differential image.
Between the two epochs, the position angle varies by $\sim 2$ degrees, and the difference in FWHM of the PSF is $<5\, \%$.

To assess the impact of noise signal on detection, we ran the same source-finding routine over the same F200W differential image of several types of galaxies (e.g., clumpy or smooth morphology with different luminosities).
Except for genuine transients such as supernovae that can be also detected in other bands, we do not detect any significant bad pixels that only appear in the single band with similar significance. 
Therefore, we conclude that the transients detected on the Dragon arc are not data-quality artifacts but real transient events.

Using the \texttt{PSFphotometry} routine of the \texttt{photutils.psf} module, we assessed the sizes of the individually detected transient events by subtracting scaled PSFs at the positions of detected transients. The residuals appear consistent with background noise, leading us to conclude that the transients are consistent with point sources.

\section*{Contamination from strong emission lines?}
We note that strong emission lines from H{\sc ii} regions still have some probability to masquerade as transients due to the imperfect match between the F182M$+$F210M and F200W filter transmission curves.
At $z=0.725$, He\,{\sc i}\, $\lambda$1.083\,$\mu$m\ and Paschen\,$\gamma$ can elevate the flux density from F182M photometry, and Paschen\,$\beta$ may boost the F200W photometry if they are strong enough.
In the ``head'' of the Dragon arc (the least magnified multiple image of the Dragon arc seen in the most eastern part of Figure 1), three bright clumps can be tentatively seen as faint F200W (2022) excess sources with peak brightness $\sim1$\,nJy\,pix$^{-1}$ in the difference image.
However, counter images of these clumps are not selected in our sample, as they are much fainter than our selected transients. Additionally, there are no underlying compact star-forming clumps seen in continuum emission.
For these reasons, we conclude that the emission line contamination in our transient sample is negligible.

\section*{Contamination from star clusters?}

We note that star clusters rather than single stars might be able to contaminate the sample if microlenses can magnify such spatially more extended objects. However, as we argue below, this is not the case and we securely reject the possibility of contamination in our microlensed transient samples from magnified star clusters.

In order to achieve a microlensing magnification boost factor of $\mu_\star = 10$ through a solar-mass lens in Abell 370 cluster given a macrolensing magnification of $100$ (Table 1), the maximum allowed source size is $r_\star = 2 \theta_E * D_s / \mu_\star$, where $\theta_E$ is the Einstein radius for the effective microlens and $D_s$ is the angular diameter distance at $z_s = 0.725$ \cite{Dai2021statsML,2008arXiv0811.0441M}. This is found to be $r_\star \sim 5.4\times10^3 (10 / \mu_\star)$\,AU.
This distance is too small compared to the typical distance between bright stars in a star cluster that can be detectable through $\sim1000 \times $ of magnifications from microlensing events\cite{2009ApJ...699..254H,2013ApJ...772...82H}.
Thus, the microlens does not provide enough magnification to such a spatially extended object.
This means that even if the observed supergiants reside in star clusters, the detected sources cannot be the entire star cluster.
Assuming an extreme case of $1000$ very bright stars within 1 pc\cite{2009ApJ...699..254H}, the projected separation between them is large ($\sim10000\,{\rm AU}$) and simultaneous microlensing boost of many of them is impossible.

\section*{Photometry}

To measure the F200W fluxes of detected transients, we performed forced aperture photometry on the difference images at the detected positions. We used a fixed aperture photometry with an aperture radius of $0.06\,{\rm arcsec}$. 
Photometric errors are estimated by placing random apertures with the same size on the transient-free regions of differential images.
To take into account realistic Poisson noise from the host galaxy, we only place random apertures where the surface brightness in F200W image is 0.4--50$\,{\rm nJy}\,{\rm pixel}^{-1}$, i.e., similar to the surface brightness of the background the Dragon arc around the transient positions.
We performed aperture correction based on the PSF models obtained using \texttt{WebbPSF} \cite{Perrin2014}.
The $5\sigma$ detection limit is therefore determined as 28.75\,AB mag in the difference image.

We also performed photometry of transients using F410M images because F410M is the only repeated filter used in both epochs while the F410M image is shallower than F200W image as it is observed as a part of grism observation in GO-3538.
During this long wavelength photometry, we limited our sample only to F410M bright sources, namely F410M detection of $>5\,\sigma$.
Because of the larger PSF size and higher sky background, longer wavelength images only have limited detection capability. Especially, within the bright region of the Dragon arc, large correlated noise buries weak transient signals. 
Additionally, larger PSF size of F410M (FWHM $\sim0^{\prime\prime}.14$) prevent us from detecting transients in crowded regions where F200W based detections have separations of $\sim0^{\prime\prime}.1$.
Therefore, we only focused on F410M photometry and analysis for the brightest microlensing events (see Supplementary Figure 1). For these bright sources, we similarly performed forced aperture photometry with a fixed aperture radius of $0.1\,{\rm arcsec}$.
The F410M detection limit of transient is $\sim27.7$\,AB\,mag.
The much smaller number of transient detections in F410M reflects the shallowness of F410M differential image, its larger PSF size, actual surface temperature of each individual star (see Figure 3), and difference of the dust opacity in each line of sight.

Photometry of microlensed stars using other filters was not performed in this study except for {\it event $\alpha$}, {\it event $\beta$}, and {\it event $\beta^{\prime}$} see \S\ref{sec:individual}.
These three events appear around the region where backgrounds are smooth and thus easy to estimate using the surrounding area. However, for most cases, the background is part of the Dragon arc itself, leading to uncertainty in estimating background and accurate photometry of each source.
This is because current observations do not have repeat observations of all filters. A full reliable SED analysis would require another imaging visit using  filters that have existing images.

\section*{Summary of transient events in 2022 and 2023}

From two independent source detections, we found 44 and 31 sources from \texttt{SExtractor} and \texttt{DAOFIND}, respectively. All \texttt{DAOFIND} detected sources are detected by \texttt{SExtractor}.
In total, we detected 45 microlensing events from the Dragon; 28 in the 2022 observation and 18 in the 2023 observation (Supplementary Table 1).
Compared with previous identifications of microlensing events and magnified single star observations (e.g., 8 in the Dragon arc identified with HST \textit{Flashlights} program down to $3.3\sigma$ significance\cite{2022arXiv221102670K}), our finding of 46 microlensing events in a single galaxy far surpasses previous records, which shows the extremely high occurrence rate of microlensing events in the Dragon arc captured by {\it JWST}.
This is because the Dragon arc has a lower redshift than other galaxies that host microlensing events e.g.,\cite{Kelly2018,Welch2022,2023A&A...672A...3D}, and therefore more stars are visible to the {\it JWST} when highly magnified.

The detection of microlensing events requires extremely high magnification ($\mu \gtrsim 10^3$). Therefore, they occur frequently around the critical curves of the large-scale gravitational lens powered by major dark matter halos of the galaxy cluster (macrolens).
The locations of our detections of microlensing events are consistent with macrolens critical curves for $z=0.725$ objects previously estimated in different studies and with different software and methods (see Figure 2).
The estimated magnifications from macrolens based on existing cluster mass models are typically at $\mu \gtrsim 10^2$ despite large uncertainties as indicated by the model-to-model scattering in Supplementary Table 1.

Although the positions of detected microlensing events are consistent with the macrolens critical curves, we note that existing lens models of Abell 370 cluster do not simultaneously explain all the microlensing events.
In particular, several transient events have large angular offsets from the expected positions of macrolens critical curves (e.g., region C in Figure 2).
These large offsets between the macrolens critical curve and microlensing events may indicate the complex structure of the dark matter sub-halo structure of the lensing cluster e.g., \cite{2017MNRAS.472.3177M,2018ApJ...867...24D, Dai2020s1226}, or uncertainty of gravitational lens models.
In-depth analyses of detailed mass models with complex sub-halo distributions incorporating the uncertainty of lens model will be important in future works.

Furthermore, we identified an event which has a large angular offset ($d_{\perp}\sim1.8^{\prime\prime}$ or $1.7\,{\rm kpc}$ assuming $z=0.735$) from the Dragon arc.
We call this event as {\it event $\alpha$} (see Figure 2), and detailed analysis and discussion is presented in Section \ref{sec:eventA}. 

We note that many point-like residuals in the Dragon arc are currently not selected as transients because of our conservative $5\sigma$ detection limit (Figure 2). These surface brightness fluctuations are also likely caused by fainter microlensed events, or events off from the peak phase.
As the magnitudes get fainter, we expect it increasingly probable that multiple, independently micro-lensed stars blended within the PSF collectively contribute to the observed flux variability. Such signals have been discussed in the contexts of pixel variability~\cite{Tuntsov2004pixelML} and collective variability of unresolved stellar associations~\cite{Dai2021statsML}.
Future deep and higher spatial resolution multi-epoch imaging observations, such as using $30\,{\rm m}$ class telescopes with the adaptive optics, will be able to resolve and identify such faint and crowded single stars.

We find that, in the region A of Figure 2, one of our sample's spatial location overlaps with the previous transient findings using HST\cite{2022arXiv221102670K}, which may represent the close microlensing events or repeating microlensing event of a same star. Although the probability of finding the same star microlensed in different epochs would be very low, it is still possible especially if the background object is a binary star system as is discussed for the finding of a $z\sim6$ star\cite{Welch2022}. Disentangling these possibilities or confirmation of such repeating events, at least, requires multi-wavelength observations of the events with the future repeat observations using {\it JWST}.
In addition, the probability of seeing multiple microlensing events from the same background star can be increased if a millilens is near the line of sight as discussed in \cite{2024arXiv240408033D}.

\section*{Color magnitude diagram}
The color magnitude diagram of the detected sources shows that the lensed sources have red colors in F200W -- F410M $> 0\,{\rm mag}$ (roughly corresponding to $J - K$ in the rest frame; Figure 3). Previous \textit{Herschel} observations at 100--500\,$\mu$m\ have shown that the Dragon arc is a dusty star-forming galaxy \cite{rawle16}, and it exhibits significant interstellar dust attenuation of up to ${\rm E(B-V)}=0.4\,{\rm mag}$ \cite{Patricio2018}. This implies a necessity of color corrections between F200W and F410M of up to $0.4\,{\rm mag}$ to derive intrinsic F200W -- F410M colors of each source, assuming a uniform dust screen.
We simply applied a color correction of $0.4\,{\rm mag}$ to demonstrate possible stellar type classification for detected sources, except for the off-arc {\it event $\alpha$} and $\beta$, but we caution the potential variation of dust attenuation of observed stars, which cannot be corrected with the data taken so far.

Sources detected in microlensing events have F200W -- F410M colors of $-0.5 \sim +0.5\,{\rm mag}$ (Figure 3). 
By comparing with theoretical stellar atmosphere spectra of giant stars (lowest surface gravity available at each effective temperature in the set of \cite{Lejeune1997}, under the assumption of $[\mathrm{M}/\mathrm{H}]=0$), these colors are consistent with typical ranges of stellar spectra with temperatures of $T\lesssim5000\,{\rm K}$.
Due to their low temperature and high luminosity, these sources are red giants/supergiants (and potentially binaries) magnified by factors of $\mu>1000$, and thus reaching absolute J-band magnitude of $-14 \sim -16$\,AB mag (see also \S\ref{sec:individual} below).
These color distributions are particularly biased toward red sources because we performed source detection in F200W filters and because of the shallowness of F410M observations. Detecting higher temperature stars at this redshift (e.g., $\gtrsim8000\,{\rm K}$) requires multi-epoch observations at $\lambda_{\rm obs}<1\,{\rm \mu m}$ \cite{2022arXiv221102670K}.

\section*{SEDs analysis of bright sources}
\label{sec:individual}

For three particularly bright events, {\it event $\alpha$}, {\it event $\beta$} and {\it event $\beta^{\prime}$}, we study their SEDs using multi-wavelength photometry.
As these events occurred in regions that are away from the brightest part of the arc, more accurate background subtraction is possible using single-epoch data.
We find full SED analysis of other transient events are difficult. This is because the proper background subtractions from their photometry are only possible for the repeating F200W and F410M bands. For {\it event $\alpha$}, {\it event $\beta$} and {\it event $\beta^{\prime}$}, backgrounds of non-repeating bands would be possible as their background is smooth. Nevertheless, their photometry might still be contaminated by their backgrounds. Thus, photometry in Supplementary Table 1 is more accurate as it uses actual backgrounds for each transient event although single epoch data in Supplementary Table 2 is useful for full SED analyses.
We conducted photometry of these events using background subtracted images of observations in 2022 and 2023.
Aperture photometry was performed with aperture radii of $0.06\,{\rm arcsec}$ and  $0.10\,{\rm arcsec}$ for short wavelength filter images and long wavelength filter images, respectively.
Annuli with a width of $0.10\,{\rm arcsec}$ were used to measure local backgrounds, and flux density uncertainties were measured from random aperture experiments. Obtained photometry is summarized in Supplementary Table 2.

\subsection*{Event $\alpha$: }
\label{sec:eventA}

From our analysis of the difference images between the 2022 and 2023 observation, {\it event $\alpha$} has the largest angular separation from the arc ($\sim1^{\prime\prime}.8$; Figure 2). 
The SED of {\it event $\alpha$} is shown in the top-left panel of Supplementary Figure 2, and we find that a $T=2200\,{\rm K}$ template among the examples in the stellar spectral library of  \cite{Lejeune1997} provides the best match to the observed SED.
Therefore, if {\it event $\alpha$} is a microlensing event of a single star, it should be a cool star in the galaxy halo of the Dragon arc. 
We also examine the F182M and F210M light curve of {\it event $\alpha$} over epoch-2 (3.4-day span in the observer frame).
Despite the data being noisy per each telescope visit, we detect no evidence of significant brightness change ($<0.7$\,mag) within this time frame. 

However, most of the existing cluster mass models do not explain the required high magnification of {\it event $\alpha$} needed to be observed as a microlensing event (typical macrolens $\mu\sim$10--100 in Supplementary Table 1). Assuming a macro$+$micro lensing magnification of $\mu=1000$ at $z_s=0.725$ and no dust extinction, {\it event $\alpha$} should have absolute $J$-band magnitude of $-7.2$\,AB mag in the rest frame.
Given the expected low number density of very luminous, short-living stars in galaxy halos, the likelihood of {\it event $\alpha$} being a microlensing supergiant star in the Dragon arc is low.
Although one can assume a significantly larger lensing magnification to accommodate a lower intrinsic luminosity of the source, this would require a much smaller impact parameter (and thus a rarer chance) for the alignment of the microlens with the background star.
{{In addition, the separation of {\it event $\alpha$} ($d_{\perp}\sim1^{\prime\prime}.8$) from the host galaxy is too large for an evolved star to be found at this distance, making the interpretation of {\it event $\alpha$} as a microlensing event even less likely.}
Therefore, we also consider the following physical explanations of {\it event $\alpha$} and explore their likelihood: 

-- Type Ia Supernova in Abell 370: \textit{Possible}. Given its proximity to one cluster member galaxy of Abell 370, {\it event $\alpha$} is likely a SN at the cluster redshift. Because most of galaxies in Abell 370 cluster are quiescent and although some cluster galaxies are known to show core-collapse supernovae e.g., \cite{Graham2012ApJ,Golubchik2023}, we consider {\it event $\alpha$} a likely SN Ia. 
At the cluster redshift $z=0.375$, the observed F200W magnitude implies an absolute magnitude of $-13.2$\,AB mag at 1.5\,$\mu$m. 
With this brightness, it can only be a nebular phase SN Ia that is $\sim200$ days post-peak in the rest frame 
 (e.g.,\cite{Graur20}).
We note that the two epochs of {\it JWST} observations have a separation of 260 days in the $z=0.375$ rest frame, placing a stringent upper limit of the age of {\it event $\alpha$} if it is a cluster SN.
The middle-left panel of Supplementary Figure 2 compares the SED of {\it event $\alpha$} with the {\it JWST} spectrum of the nearby Type Ia supernova SN 2021aefx at $+$255 days post peak (i.e., nebular phase \cite{Kwok23}).
We find that the observed SED of {\it event $\alpha$} can be explained by this spectral template with similar reduced $\chi^2_{\rm reduced}$ (of $=1.6$) to that of the aforementioned cool star template ( $\chi^2_{\rm reduced}=1.6$), and the residual is mostly dominated by the F460M non detection.
Therefore, we conclude that {\it event $\alpha$} could be a SN Ia at the redshift of the foreground cluster.

-- Kilonova in Abell 370: \textit{Unlikely}. At $z=0.375$, the observed F200W magnitude could match that of a kilonova like AT 2017gfo (the electromagnetic counterpart of GW170817\cite{ligo17}) at $\sim10$ days post merger \cite{villar17}.
At this epoch, AT 2017gfo has already shown a red rest-frame $H - K$ color of 1\,AB mag. 
However, this is not observed for {\it event $\alpha$} in F210M -- F300M.
Therefore, we conclude that {\it event $\alpha$} is unlikely a kilonova at the cluster redshift. However, our knowledge on kilonovae's light curves and SEDs could still be very limited, which could show a large diversity because of different ejecta properties \cite{kawaguchi20}.

-- Luminous red nova in Abell 370 or the Dragon arc: \textit{Unlikely}. Massive star merger can trigger luminous red novae reaching absolute V-band magnitude of $\sim-15$\,mag e.g., \cite{smith16}.
They can also show similar rest-frame near-infrared color as that of {\it event $\alpha$}. 
However, we argue that the likelihood of having a massive stellar progenitor in either the halo of Dragon or Abell 370 quiescent member galaxies is considerably low.

-- ``Hostless'' SN at $z\simeq 1 - 2$: \textit{Possible}. 
{\it Event $\alpha$} could also be a lensed SN hosted by a low-mass galaxy that was not detected even by {\it JWST}.
This is also known as ``hostless'' supernova. 
If {\it event $\alpha$} is a SN at $z\gtrsim 2$, based on \texttt{glafic} lens model \cite{Oguri2010,2018ApJ...855....4K}, we expect the detection of a counter image of the SN with a short time delay to {\it event $\alpha$} ($\lesssim20$\,days) but potentially larger magnification.
If {\it event $\alpha$} is a SN at $z\lesssim 0.8$, then with the large magnification ($\mu\sim10$) and limited evolution time ($< 1$\,yr in observed frame), {\it event $\alpha$} should have been as bright as $\lesssim27$\,AB mag at 2\,$\mu$m, i.e., 1\,mag brighter than the actual observed value.
Therefore, if {\it event $\alpha$} is a hostless SN, it can only be at $z\simeq 1 - 2$.
With the SN Ia spectral templates from \cite{hsiao07} and software \texttt{SNCosmo} \cite{sncosmo}, we find that the SED of {\it event $\alpha$} can be explained by SN Ia at $z\sim 1.32$ with $\chi^2_{\rm reduced} = 1.7$, for which the age is $t\sim +129$\,days post-peak ($+300$ days in the observer frame; Supplementary Figure 2, bottom-left panel).
At this redshift, the multiple images are expected as nondetections in obtained {\it JWST} data, either because of large difference of arrival time or proximity to bright cluster member galaxies.
Therefore, we do not rule out the possibility of {\it event $\alpha$} being a hostless SN at $z\sim1.3$.

As a final remark, naturally there are other transient sources (e.g., tidal disruption events) that could be considered to interpret {\it event $\alpha$}. However, we believe that the above discussions cover most of the known possibilities, and we conclude that {\it event $\alpha$} could be also interpreted as a nebular-phase SN Ia at Abell 370 redshift or a ``hostless'' SN at $z\sim1.3$.

\subsection*{Event $\beta$ and $\beta^{\prime}$: }
The {\it event $\beta$} and {\it event $\beta^{\prime}$} are found in the 2022 and 2023 observations, respectively, at the most south-east position of the Dragon arc (Figure 2).
Both {\it event $\beta$} and {\it event $\beta^{\prime}$} are some of the brightest events among the current microlensing events found in the 2022 and 2023 images.
As the locations of these events are in the outer disk region of the Dragon arc, we expect low dust attenuation for these two sources.
Thus, we assume no dust attenuation for sources observed in both {\it event $\beta$} and  {\it event $\beta^{\prime}$}. 

Right panel of Supplementary Figure 2 shows the observed fluxes of {\it event $\beta$} and {\it event $\beta^{\prime}$}. Using the spectra of stellar atmosphere models of \cite{Lejeune1997}, we performed $\chi^{2}$ minimization.
We find that low temperature stars with surface temperature of $T=3000\,{\rm K}$ to $T=3300\,{\rm K}$ fit well and a $T=3200\,{\rm K}$ star gives the minimum $\chi^{2}_{\rm reduced}=1.1$ for the SED of {\it event $\beta$}. Similarly, the observed SED of {\it event $\beta^{\prime}$} is well represented with a star with surface temperature of $T=3500 - 3700\,{\rm K}$ while a $T=3500\,{\rm K}$ star gives the minimum $\chi^{2}_{\rm reduced}=4.7$.
With these fits, we find the apparent luminosity of the  {\it event $\beta$} and  {\it event $\beta^{\prime}$} to be $\mu\,L_{\rm bol} \sim 1\times10^{8}\,L_{\odot}$ and $\mu\,L_{\rm bol} \sim 8\times10^{7}\,L_{\odot}$, respectively.

Due to their low temperatures represented by their red colors, these events would be microlensed red giant/supergiant stars that are strongly magnified with $\mu \gtrsim 1000$.
The different spectral shapes of {\it event $\beta$} and {\it event $\beta^{\prime}$} suggest that they would be different stars, rather than multiple images of a single star.
At the same time, as low temperature stars at high-redshift are only bright in $\lambda_{\rm obs}\gtrsim2\,{\rm \mu m}$, these sources demonstrate the importance of such near-IR observations to build complete stellar samples at high-redshift.

An alternative interpretation would be that these two stars are indeed the same star multiply imaged as a result of astrometric perturbation induced by dark matter subhalos. Such events are predicted by \cite{2018ApJ...867...24D} and will occur specifically around the critical crossing region such as the location near {\it event $\beta$} and $\it event {\beta^{\prime}}$ see also \cite{Kelly2018}. In this case, the difference of the observed SEDs are interpreted by intrinsic variability of the magnified star. In fact, such variability of stars accompanying changes of surface temperature and SEDs are observed in nearby stars such as Cepheid. Although a definitive conclusion is difficult to obtain with existing photometry using single epochs, confirmations of such magnified close-paired star in large numbers allow to investigate the subhalo structure of dark matter as well as to constrain dark matter physics. Future {\it JWST}'s time-domain observations using matched filters allow us to obtain accurate multi-wavelength photometry of individually magnified stars and allow us to conduct such studies.

\bmhead{Data availability}
The $2\,{\rm \mu m}$ differential image generated and analysed during the current study is available from \url{https://github.com/yfudamoto/the_dragon_arc2024.git}.
Other datasets generated are available from the corresponding author on reasonable request.
The raw data from GTO-1208 (CANUCS) are available on the Mikulski Archive for Space Telescopes (MAST) at doi:10.17909/ph4n-6n76 \cite{MASTDOI}.

\bmhead{Acknowledgements}
We acknowledge reviewers who provided us constructive comments and suggestions based on their expertise that significantly improved our paper.
We thank Lindsey Kwok for kindly sharing the {\it JWST} spectrum of SN 2021aefx, and Jordi Miralda-Escude for insightful comments.
This work is based on observations made with the NASA/ESA/CSA James Webb Space Telescope. The data were obtained from the Mikulski Archive for Space Telescopes at the Space Telescope Science Institute, which is operated by the Association of Universities for Research in Astronomy, Inc., under NASA contract NAS 5-03127 for {\it JWST}. These observations are associated with program \#1208, 3538.
The authors acknowledge the {\it JWST} GO-3538 team led by PI E.\ Iani for developing their observing program with a zero-exclusive-access period.
This work was supported by JSPS KAKENHI Grant Number JP22K21349, JP23K13149, JP20H05856, JP22H01260, JP22J21.
{F.S., E.E., and Y.Z.} acknowledge JWST/NIRCam contract to the University of Arizona NAS5-02015.
F.S. acknowledges support for program \#2883 provided by NASA through a grant from the Space Telescope Science Institute, which is operated by the Association of Universities for Research in Astronomy, Inc., under NASA contract NAS 5-03127. 
J.M.D. acknowledges the support of project PID2022-138896NB-C51 (MCIU/AEI/MINECO/FEDER, UE) Ministerio de Ciencia, Investigaci\'on y Universidades.
L.D. acknowledges research grant support from the Alfred P. Sloan Foundation (Award Number FG-2021-16495), and support of Frank and Karen Dabby STEM Fund in the Society of Hellman Fellows.
A.Z. acknowledges support by Grant No. 2020750 from the United States-Israel Binational Science Foundation (BSF) and Grant No. 2109066 from the United States National Science Foundation (NSF); by the Ministry of Science \& Technology, Israel; and by the Israel Science Foundation Grant No. 864/23.
E.Z. acknowledges project grant 2022-03804 from the Swedish Research Council (Vetenskapsr\aa{}det) and has also benefited from a sabbatical at the Swedish Collegium for Advanced Study.
M.J. is supported by the United Kingdom Research and Innovation (UKRI) Future Leaders Fellowship `Using Cosmic Beasts to uncover the Nature of Dark Matter' (grants number MR/S017216/1 \& MR/X006069/1).
D.J.L. acknowledges support from the UKRI FLF grants number MR/S017216/1 \& MR/X006069/1.
D.J.L. is also supported by Science and Technology Facilities Council (STFC) grants ST/T000244/1 and ST/W002612/1.
E.I. acknowledges funding from the Netherlands Research School for Astronomy (NOVA). 
R.A.W. and S.H.C. acknowledge support from NASA JWST Interdisciplinary Scientist grants NAG5-12460, NNX14AN10G and 80NSSC18K0200 from GSFC.
C.-C.C. acknowledges support from the National Science and Technology Council of Taiwan (111-2112M-001-045-MY3), as well as Academia Sinica through the Career Development Award (AS-CDA-112-M02).
D.E. acknowledges support from a Beatriz Galindo senior fellowship (BG20/00224) from the Spanish Ministry of Science and Innovation, projects PID2020-114414GB-100 and PID2020-113689GB-I00 financed by MCIN/AEI/10.13039/501100011033, project P20-00334  financed by the Junta de Andaluc\'{i}a, and project A-FQM-510-UGR20 of the FEDER/Junta de Andaluc\'{i}a-Consejer\'{i}a de Transformaci\'{o}n Econ\'{o}mica, Industria, {Conocimiento y}, Universidades.
K.K. acknowledges the support by JSPS KAKENHI Grant Number JP17H06130, JP22H04939, JP23K20035, {and JP24H00004}.
G.E.M. acknowledges financial support from the Villum Young Investigator grants 37440 and 13160 and the Cosmic Dawn Center (DAWN), funded by the Danish National Research Foundation (DNRF) under grant No. 140.
A.N. acknowledges funding from Olle Engkvists Stiftelse.
F.E.B. acknowledges support from ANID-Chile BASAL CATA FB210003, FONDECYT Regular 1241005,
and Millennium Science Initiative Program  {AIM23-0001} and ICN12\_009. P.L.K acknowledges grant AAG 2308051 from the NSF.

\bmhead{Author Contributions}
Y.F. wrote the main part of the text, analyzed the data, produced figures.
F.S. calibrated all data, analyzed the data and contributed text. J.M.D., M.O., A.Z., M.J., D.J.L, A.K., H.K. contributed analysis and interpretations of the results. E.Z. contributed SED fitting of detected stars. E.E. and R.W. contributed interpretations of the data. E.I. contributed to the planning and execution of the JWST GO-3538 programs. All co-authors contributed to the scientific interpretation of the results, and helped to write up the manuscript.

\bmhead{Competing interests}
The authors declare no competing interests.
Correspondence and requests for materials should be addressed to Y.F.~(email: yoshinobu.fudamoto@gmail.com).

\break
\begin{table*}[!h]
    \renewcommand{\arraystretch}{1.1}
    \footnotesize
    \centering
    \caption{F200W-based detections of microlensing events in 2022 and 2023}
    \begin{tabular}{lllclccccc}
    \hline
        ID & RA & Dec & year & mag$_{\rm F200W}$ & $\mu$ glafic$^{\rm a}$   & $\mu$ WSLAP+$^{\rm b}$  & $\mu$ Zitrin$^{\rm c}$ & $\mu$ BUFFALO$^{\rm d}$ & $\mu$ HFF$^{\rm e}$\\
    \hline2201 & 39.969401 & -1.584803 & 2022 & $28.4 \pm 0.1$      & $26^{+9}_{-6}$      & $65$      & $88$      & $2307$      & $37$  \\
2202 & 39.969417 & -1.584746 & 2022 & $28.6 \pm 0.2$      & $24^{+7}_{-5}$      & $34$      & $78$      & $317$      & $27$  \\
2203 & 39.969426 & -1.58478 & 2022 & $28.3 \pm 0.1$      & $33^{+12}_{-8}$      & $47$      & $87$      & $2348$      & $44$  \\
2204 & 39.969442 & -1.584832 & 2022 & $28.3 \pm 0.1$      & $130^{+264}_{-61}$      & $357$      & $185$      & $46$      & $890$  \\
2205 & 39.969469 & -1.584738 & 2022 & $27.6 \pm 0.1$      & $43^{+18}_{-11}$      & $50$      & $92$      & $2575$      & $46$  \\
2206 & 39.969659 & -1.584863 & 2022 & $28.5 \pm 0.2$      & $27^{+5}_{-4}$      & $46$      & $13665$      & $31$      & $31$  \\
2207 & 39.969942 & -1.584886 & 2022 & $28.3 \pm 0.1$      & $30^{+5}_{-4}$      & $102$      & $168$      & $56$      & $50$  \\
2208 & 39.970067 & -1.584966 & 2022 & $28.6 \pm 0.2$      & $29^{+4}_{-3}$      & $123$      & $180$      & $55$      & $50$  \\
2209 & 39.970094 & -1.585033 & 2022 & $28.6 \pm 0.2$      & $26^{+3}_{-3}$      & $73$      & $385$      & $44$      & $43$  \\
2210 & 39.970136 & -1.584935 & 2022 & $28.2 \pm 0.1$      & $37^{+7}_{-5}$      & $4639$      & $108$      & $86$      & $76$  \\
2211 & 39.970328 & -1.585 & 2022 & $28.6 \pm 0.2$      & $48^{+14}_{-8}$      & $134$      & $91$      & $371$      & $333$  \\
2212 & 39.970374 & -1.585049 & 2022 & $28.7 \pm 0.2$      & $46^{+13}_{-8}$      & $117$      & $100$      & $163$      & $721$  \\
2213 & 39.970374 & -1.585009 & 2022 & $27.9 \pm 0.1$      & $54^{+18}_{-10}$      & $88$      & $82$      & $540$      & $1452$  \\
2214 & 39.9704 & -1.585094 & 2022 & $27.5 \pm 0.1$      & $43^{+11}_{-7}$      & $182$      & $108$      & $123$      & $355$  \\
2215 & 39.970419 & -1.585006 & 2022 & $28.1 \pm 0.1$      & $64^{+28}_{-15}$      & $65$      & $74$      & $1733$      & $3059$  \\
2216 & 39.970445 & -1.584967 & 2022 & $28.2 \pm 0.2$      & $89^{+66}_{-26}$      & $51$      & $65$      & $1795$      & $313$  \\
2217 & 39.970468 & -1.585274 & 2022 & $28.3 \pm 0.1$      & $32^{+5}_{-4}$      & $161$      & $691$      & $54$      & $95$  \\
2218 & 39.970727 & -1.585275 & 2022 & $28.5 \pm 0.1$      & $60^{+24}_{-13}$      & $88$      & $146$      & $1935$      & $110$  \\
2219 & 39.971079 & -1.584887 & 2022 & $28.4 \pm 0.1$      & $29^{+10}_{-5}$      & $15$      & $34$      & $15$      & $10$  \\
2220 & 39.971768 & -1.584963 & 2022 & $28.5 \pm 0.2$      & $196^{+443}_{-99}$      & $114$      & $182$      & $65$      & $3495$  \\
2221 & 39.971904 & -1.584807 & 2022 & $28.5 \pm 0.2$      & $116^{+199}_{-47}$      & $119$      & $140$      & $48$      & $207$  \\
2222 & 39.971952 & -1.584836 & 2022 & $27.3 \pm 0.1$      & $94^{+102}_{-33}$      & $116$      & $166$      & $62$      & $56$  \\
2223 & 39.972045 & -1.584789 & 2022 & $27.4 \pm 0.1$      & $71^{+41}_{-20}$      & $156$      & $161$      & $98$      & $28$  \\
2224 & 39.97209 & -1.584741 & 2022 & $28.4 \pm 0.1$      & $73^{+43}_{-21}$      & $568$      & $155$      & $112$      & $26$  \\
2225 & 39.972163 & -1.584835 & 2022 & $27.5 \pm 0.1$      & $29^{+6}_{-4}$      & $45$      & $120$      & $39$      & $8$  \\
2226 & 39.972248 & -1.584707 & 2022 & $27.3 \pm 0.0$      & $25^{+5}_{-4}$      & $30$      & $104$      & $26$      & $7$  \\
2227$^{\ast}$ & 39.972783 & -1.584734 & 2022 & $27.4 \pm 0.0$      & $30^{+10}_{-6}$      & $418$      & $456$      & $22$      & $11$  \\
2301 & 39.969121 & -1.5846 & 2023 & $28.6 \pm 0.2$      & $18^{+7}_{-5}$      & $18$      & $119$      & $363$      & $18$  \\
2302$^{\dagger}$ & 39.969252 & -1.584188 & 2023 & $28.0 \pm 0.1$      & $11^{+3}_{-3}$      & $11$      & $189$      & $16$      & $8$  \\
2303 & 39.969456 & -1.584804 & 2023 & $28.5 \pm 0.2$      & $85^{+97}_{-32}$      & $191$      & $139$      & $73$      & $913$  \\
2304 & 39.969473 & -1.584769 & 2023 & $28.5 \pm 0.2$      & $67^{+48}_{-21}$      & $84$      & $113$      & $430$      & $96$  \\
2305 & 39.970074 & -1.584921 & 2023 & $27.2 \pm 0.1$      & $34^{+6}_{-4}$      & $286$      & $122$      & $73$      & $64$  \\
2306 & 39.970191 & -1.585072 & 2023 & $28.6 \pm 0.2$      & $29^{+4}_{-3}$      & $114$      & $249$      & $51$      & $52$  \\
2307 & 39.970362 & -1.585098 & 2023 & $28.1 \pm 0.2$      & $39^{+8}_{-6}$      & $357$      & $126$      & $90$      & $134$  \\
2308 & 39.970436 & -1.585247 & 2023 & $28.1 \pm 0.1$      & $32^{+5}_{-4}$      & $193$      & $405$      & $55$      & $89$  \\
2309 & 39.970446 & -1.585053 & 2023 & $27.8 \pm 0.2$      & $58^{+22}_{-12}$      & $79$      & $85$      & $1241$      & $2646$  \\
2310 & 39.970578 & -1.585187 & 2023 & $28.7 \pm 0.2$      & $55^{+19}_{-11}$      & $95$      & $114$      & $648$      & $250$  \\
2311 & 39.970619 & -1.585151 & 2023 & $27.7 \pm 0.1$      & $78^{+46}_{-21}$      & $54$      & $83$      & $2048$      & $85$  \\
2312 & 39.971437 & -1.584837 & 2023 & $28.4 \pm 0.1$      & $151^{+331}_{-72}$      & $48$      & $66$      & $110$      & $36$  \\
2313 & 39.971446 & -1.584912 & 2023 & $28.2 \pm 0.1$      & $173^{+382}_{-85}$      & $51$      & $75$      & $231$      & $51$  \\
2314 & 39.971971 & -1.584861 & 2023 & $28.6 \pm 0.2$      & $85^{+72}_{-27}$      & $111$      & $176$      & $80$      & $40$  \\
2315 & 39.972012 & -1.584799 & 2023 & $27.5 \pm 0.1$      & $79^{+59}_{-24}$      & $146$      & $162$      & $73$      & $36$  \\
2316 & 39.972014 & -1.584859 & 2023 & $28.5 \pm 0.2$      & $69^{+40}_{-19}$      & $104$      & $177$      & $140$      & $27$  \\
2317 & 39.972169 & -1.584785 & 2023 & $28.1 \pm 0.1$      & $32^{+7}_{-5}$      & $53$      & $134$      & $51$      & $9$  \\
2318$^{\ast\ast}$ & 39.972768 & -1.584761 & 2023 & $26.8 \pm 0.0$      & $24^{+7}_{-5}$      & $105$      & $178$      & $17$      & $8$  \\
    \hline
    \end{tabular}
    Noteworthy events include $\dagger$: {\it event $\alpha$}, $\ast$: {\it event $\beta^{\prime}$} and $\ast\ast$: {\it event $\beta$} (see discussions in \S\ref{sec:eventA}). Lensing magnifications are from ${\rm a}$: \cite{2018ApJ...855....4K}; ${\rm b}$: \cite{Diego2007}; ${\rm c}$: \cite{Zitrin2009}; ${\rm d}$: \cite{2023MNRAS.524.2883N}; ${\rm e}$: \cite{2022MNRAS.514..497L}.
    Note: Near the critical curve, uncertainties of magnifications are large and dominated by systematic of each lens model \cite{Meneghetti2017}. Although the uncertainties of glafic model are displayed to demonstrate statistical uncertainties, intrinsic uncertainties are not captured by statistical errors of each model. Thus, we list predicted magnification from different models to indicate typical uncertainties.
\label{tab:events}
\end{table*}

\break
\begin{table*}[!h]
    \renewcommand{\arraystretch}{0.9}
    \centering
    \caption{Multi-wavelength Photometry of {\it Evant} $\alpha$, $\beta$, and $\beta^{\prime}$}
    \begin{tabular}{c|cc||c|c}
    \hline
     Filter set 1 & {\it Event} $\alpha$ & {\it Event} $\beta^{\prime}$ & Filter set 2 & {\it Event} $\beta$\\
    \hline
    F182M & $28.0\pm0.3$ & $26.7\pm0.1$ & F090W & $30.1\pm0.7$\\
    F210M & $27.8\pm0.3$ & $26.5\pm0.1$ & F115W & $28.7\pm0.3$\\
    F300M & $27.9\pm0.5$ & $25.9\pm0.2$ & F150W & $27.6\pm0.1$\\
    F335M & $28.2\pm0.5$ & $26.0\pm0.2$ & F200W & $27.0\pm0.2$ \\
    F410M & $27.9\pm0.3$ & $26.3\pm0.2$ & F277W & $26.5\pm0.2$\\
    F460M & $>27.5$ & $26.6\pm0.3$ & F356M & $26.5\pm0.2$ \\
     & & & F410M & $26.9\pm0.2$\\
     & & & F444W & $27.0\pm0.2$\\
    \hline
    \end{tabular}
    \\
    All photometry are in AB magnitude
    \label{tab:individual}
\end{table*}

\break
\begin{figure}[!h]
\centering
\includegraphics[width=0.9\textwidth]{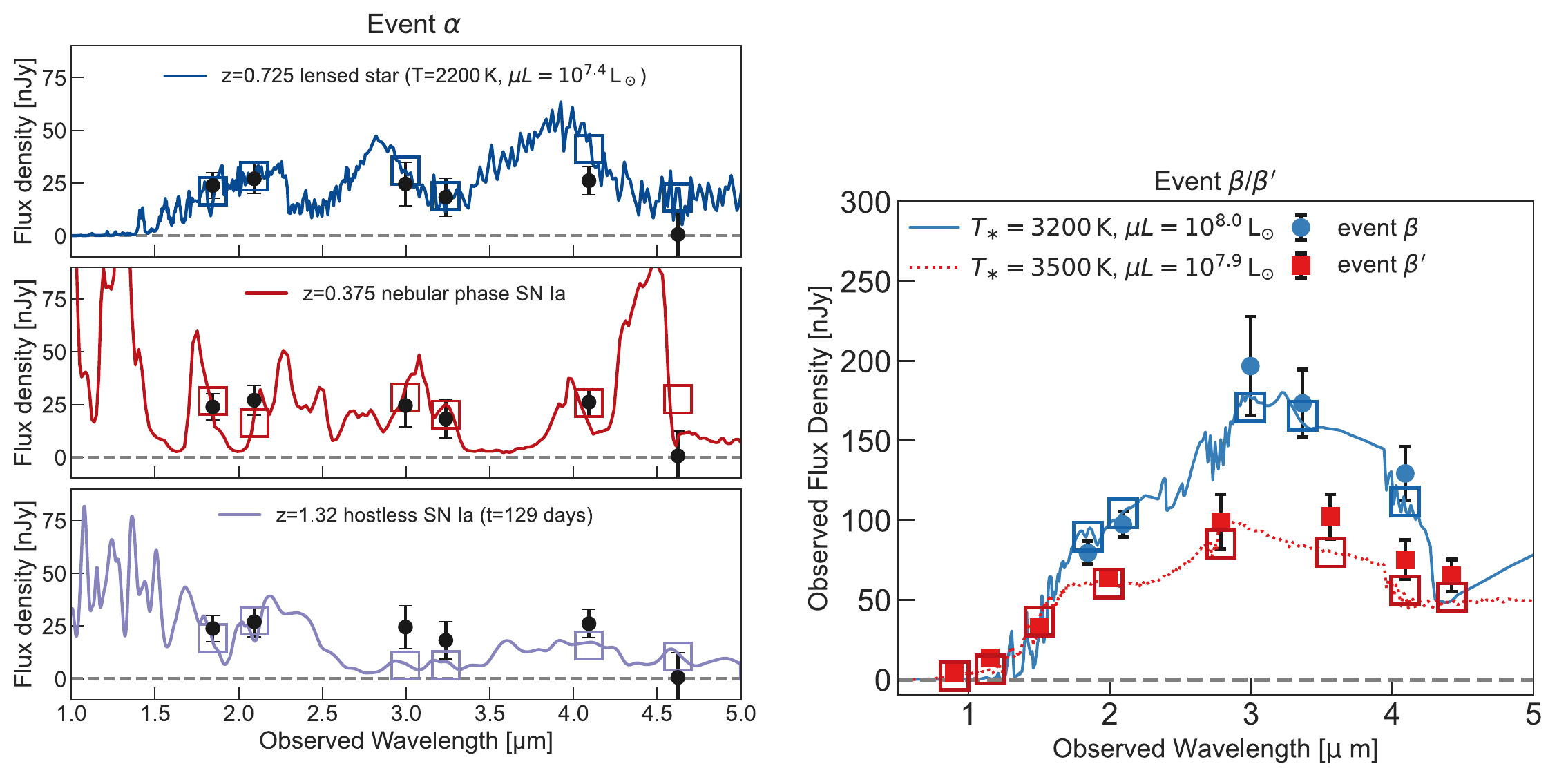}
\caption{JWST single-epoch SEDs of {\it event} $\alpha$ (left) and event $\beta$, $\beta^\prime$ (right).
For {\it event} $\alpha$, we consider three possible scenarios, including (\romannumeral1) microlensed red supergiant star in the halo of the Dragon arc (model from \cite{Lejeune1997}; top pannel); (\romannumeral2) nebular-phase SN Ia in the foreground Abell 370 cluster (spectrum from \cite{Kwok23}; middle panel); (\romannumeral3) lensed SN Ia at $z=1.32$ with no obvious host detection (129-day post-peak, \cite{hsiao07}; bottom panel).
All three templates fit the observed SED with similar goodness.
For event $\beta$ / $\beta^{\prime}$, although they are found at close positions ($\Delta D=0^{\prime\prime}.1$) with similar F200W fluxes, their SEDs are largely different, showing that {\it event $\beta$} and {\it event $\beta^{\prime}$} are different objects experiencing different microlensing events.
}
\label{fig:sed}
\end{figure}

\break
\begin{figure}[!h]
    \centering
    \includegraphics[width=1.0\textwidth]{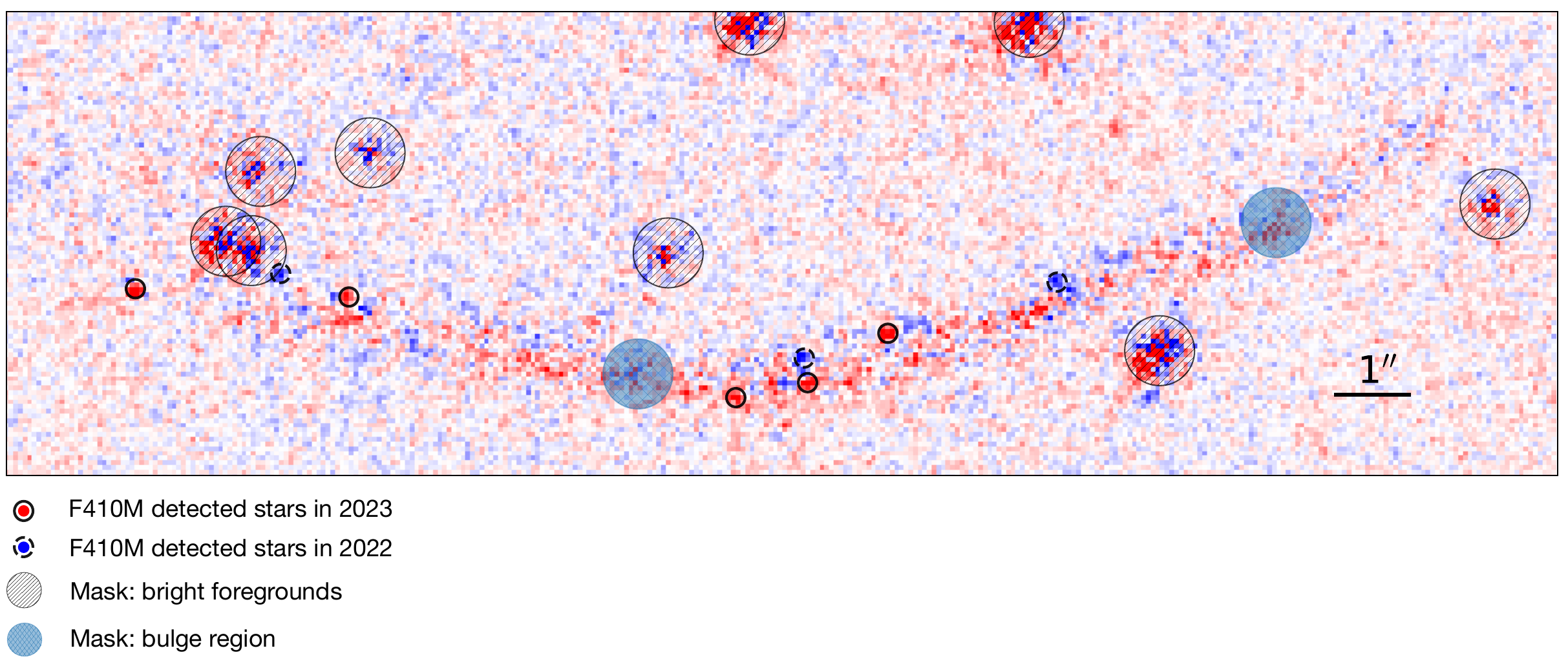}
    \caption{The observed $\sim4\,{\rm \mu m}$ differential image between the 2022 and 2023 epochs. Solid and dashed circles show locations of detected lensed stars in the F200W differential image that have $>5\,\sigma$ F410M flux measurements at the same time.
    Positive signals (red) show objects that appear only in 2023, while negative signals (blue) show sources only seen in 2022. 
    Hatched circles show masked regions to avoid contamination from bright residuals. In total, eight microlensed stars have $>5\,\sigma$ F410M flux measurements.}
    \label{fig:F410Mdiff}
\end{figure}

\pagebreak


\end{document}